# Room-temperature magnetism and controlled cation distribution in vanadium ferrite thin films

Antonio Peña Corredor[1], Matthieu Gamarde[1], Lamiae El Khabchi[1], María José Vázquez Bernárdez[1], Marc Lenertz[1], Cédric Leuvrey[1], Laurent Schlur[1], François Roulland[1], Nathalie Viart[1], Christophe Lefevre[1*]

[1]Institut de Physique et Chimie des Matériaux de Strasbourg. 23 rue du Loess, 67200 Strasbourg, France.

*corresponding author: christophe.lefevre@ipcms.unistra.fr

**Abstract**

Spinel oxides demonstrate significant technological promise due to the vast array of interrelated physical properties that their unique structure supports. Specifically, the $Fe_{1+x}V_{2-x}O_4$ spinel system garners extensive interest due to the presence of orbitally ordered states and multiferroism. This study focuses on the elaboration of high-quality $Fe_2VO_4$ (x = 1) thin films on MgO substrates via pulsed laser deposition. Structural analyses confirm the epitaxial growth of the films, their high crystallinity and fully strained nature. The cationic distribution and stoichiometry were investigated using Resonant Elastic X-ray Scattering experiments, in conjunction with comprehensive characterization of the films' physical and electrical properties. The films exhibit room-temperature magnetism, with a magnetization consistent with the $(Fe^{3+})_{Td}[Fe^{2+}V^{3+}_2]_{Oh}O_4$ inverse spinel structure unveiled by anomalous diffraction. This work represents the inaugural successful deposition of $Fe_2VO_4$ thin films, thereby expanding the family of spinel vanadium oxide thin films with a new member that demonstrates room-temperature magnetic properties.

# 1. Introduction

Spinel oxides have garnered considerable attention over time, primarily because of the exceptional versatility inherent to the $AB_2O_4$ structure. This particular structural configuration can host a broad spectrum of cations within its tetrahedral (Td) and octahedral (Oh) sites, thereby yielding a variety of orderings - magnetic, charge, and orbital, among others [1]–[4]. This makes spinel oxides strong contenders for future energy-efficient applications such as magnetoelectric memories or spintronic devices. [5]–[7]

Among all members of the spinel oxide family, the $Fe_{1+x}V_{2-x}O_4$ spinel system has the particularity of adopting either a direct or inverse spinel structure depending on the proportion of $Fe^{3+}$, $V^{3+}$ and $Fe^{2+}$ cations that the structure hosts. For low iron concentrations ($0 \leq x \leq 0.35$), and just like the $FeV_2O_4$ (FVO) iron vanadate, the system adopts a direct spinel structure: $(Fe^{2+})[Fe^{3+}_xV^{3+}_{2-x}]O_4$, For higher iron contents ($0.35 \leq x \leq 1$), the structure slowly transforms into a 3-2-3 inverse spinel structure: $(Fe^{3+})[Fe^{2+}_xV^{3+}_{2-x}]O_4$, and such is the case of the $Fe_2VO_4$ (VFO) vanadium ferrite. The inverse spinel structure is kept for higher iron concentrations ($1 \leq x \leq 2$): $(Fe^{3+})[Fe^{2+}Fe^{3+}_{x-1}V^{3+}_{2-x}]O_4$, with the trivalent iron progressively replacing the trivalent vanadium, until it reaches the composition of magnetite ($Fe_3O_4$). [8]–[10]

The three cations in $Fe_{1+x}V_{2-x}O_4$ present a magnetic behaviour, with the following spin values: $S\text{-}Fe^{3+} = 2.5$, $S\text{-}Fe^{2+} = 2$ and $S\text{-}V^{3+} = 1$. Their relative orientation, majorly ruled by antiferromagnetic super-exchange interactions between Oh- and Td-sites, determine the material's magnetism, which can adopt complex noncollinear magnetic textures. [10] Furthermore, two of the cations are Jahn-Teller (JT) active: $V^{3+}$ in an octahedral site exhibits a $t_{2g}^2$ electronic configuration and $Fe^{2+}$ either a $t_{2g}^4e_g^2$ or a $e^3t_2^3$ depending on whether it occupies an octahedral or a tetrahedral site, respectively. $V^{3+}$ also presents a non-negligible orbital moment [11] which results into spin-orbit coupling (SOC) effects, increasing the complexity of interactions in the $Fe_{1+x}V_{2-x}O_4$ system.

The interplay between JT and SOC phenomena explains why FVO (x = 0) adopts diverse orbitally-ordered states and undergoes up to four structural transformations with temperature. [12], [13] Along with these structural changes, FVO's magnetic structure changes from paramagnetic to collinear ferrimagnetic, and then to non-collinear ferrimagnetic. [11] FVO has also proved to show a ferroelectric behaviour [14], making this spinel vanadate compound a multiferroic oxide with interconnected orbital, spin and charge degrees of freedom. Despite all of these functionalities, the fact that the ferroic orderings in FVO only occur at low temperatures (below 110 K) hinders its implementation into functional devices.

One approach to move FVO's properties up in temperature is through the addition of a strain degree of freedom, for instance, depositing the material as thin films [15]. On top of that, thin films present promising avenues to utilize the coupling of ferroic order parameters in practical devices for logic, memory, and sensor technologies. [16] The deposition in thin films has however only increased the magnetic transition temperature up to 150 K [17]. The increase in iron content is another strategy for the transition temperature increase, with the Curie temperature ($T_C$) arising from 454 K for VFO to 851 K for $Fe_3O_4$. [10]

On the one hand, $Fe_3O_4$ (x = 2) presents a magnetic ordering at room temperature, with a cubic-to-monoclinic Verwey transition at 120 K. [18] In the thin-film form it displays a giant magnetoresistance behaviour whose temperature dependence is itself affected by the Verwey transition, and the material's properties as a thin film have been largely documented in the literature [19]–[21].

One the other hand, the understanding of VFO's (x = 1) is comparatively limited. In the bulk form, the structure and magnetic properties of the $Fe_{1+x}V_{2-x}O_4$ spinel system have been documented by Wakihara *et* al. [10]. VFO crystallises in a cubic spinel structure with a = 8.418(2) Å, presents room-temperature ferrrimagnetism and a saturation magnetization ($M_s$) of around 0.73 $\mu_B$ per formula unit ($\mu_B$/f.u.) at 77 K. Mössbauer spectroscopy studies have confirmed the cationic arrangement corresponding to an inverse spinel structure. [8], [22] Consistent results regarding the $M_s$ and the cationic distribution have been reported in micron-sized VFO coatings [23]. The exploration of VFO in thin films however remains until now absent from the literature.

In this report, we present the elaboration of $Fe_2VO_4$, a material in the crossroads of the more widely studied FVO and $Fe_3O_4$, in thin films by Pulsed Laser Deposition (PLD). Firstly, we have obtained dense crystalline targets of bulk VFO, and the deposition conditions have been optimised for its deposition onto MgO substrates. The deposition of VFO onto MgO is expected to happen in a "cube on four-cubes" arrangement, as schematised in **Figure 1**.

High-quality epitaxial thin films have been obtained through this method. The cationic occupation has been confirmed via anomalous diffraction methods. Besides, the transport and magnetic properties of the films have been characterised, confirming room-temperature ferrimagnetism and an inverse spinel structure. This study reports the first successful deposition of VFO thin films and enlarges the family of spinel vanadium oxide thin films with a novel, room-temperature magnetic member.

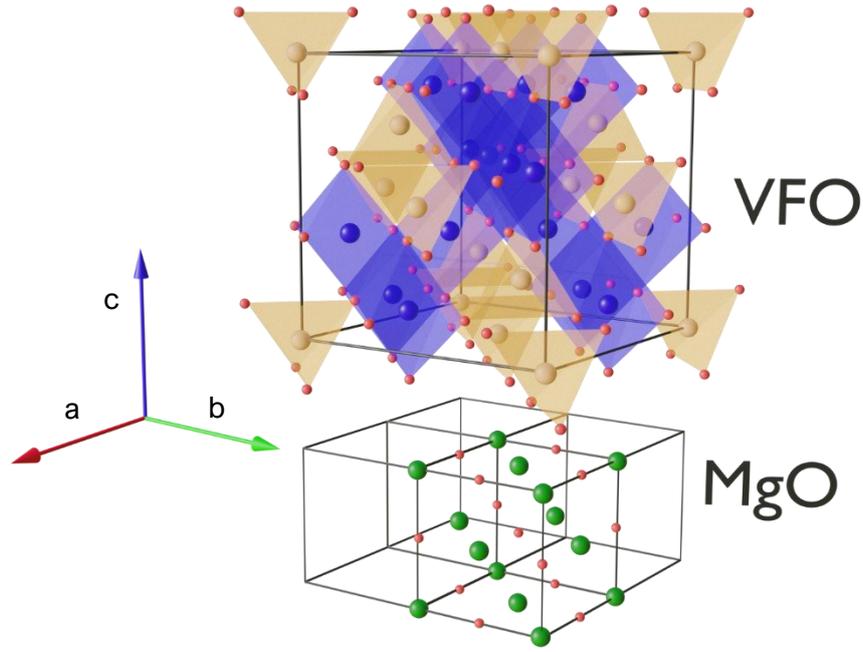

**Figure 1**. Schematic representation of epitaxial VFO (top) and MgO (bottom). On VFO: blue – cations at Oh site, orange – cations at Td site. MgO: green – Mg cations. In both: red – O.

## 2. Materials and methods

The $Fe_2VO_4$ target was prepared using a classical solid-state ceramic method, taking the route optimized for the $FeV_2O_4$ synthesis as a starting point [14], [17]. High purity commercial precursor powders of α-$Fe_2O_3$ (Strem Chemicals, 99.8%) and $V_2O_5$ (Aldrich, 99.6%) were weighed in stoichiometric conditions and mixed during 15 minutes in an agate stone mortar grinder with a small quantity of ethanol. The resulting slurry was then dried and placed on a platinum crucible which was itself placed in an alumina crucible. The precursor mixture was calcinated in a tubular furnace under a mildly reducing atmosphere of 99.5% Ar/0.5% $H_2$ at a flow of 110 sscm. It was first taken up to 600°C at a heating rate of 100°C/h and left at this temperature for ten hours. Secondly, the mix was heated up to 1100°C and kept at this temperature for ten hours, for the formation of the spinel phase. The calcinated powder was finally cooled down at a rate of 200°C/h in the same atmosphere. The overall solid-state reaction process has been summarized in **Equation (1)**.

$$Fe_2O_3 + \tfrac{1}{2} V_2O_5 \xrightarrow[\substack{i)\ 600°C,\ 10\ h \\ ii)\ 1100°C,\ 10\ h}]{99.5\%\ Ar/0.5\%\ H_2} Fe_2VO_4 \qquad (1)$$

The partial reduction of Fe(III) into Fe(II) requires a mildly reducing atmosphere, in order to avoid a complete reduction into Fe(II) or even the obtention of metallic Fe. The transformation of V(V) into V(III) requires more severe reducing conditions and it is extremely important to have V(V) reduced into V(III) before heating to temperatures higher than 690ºC, because $V_2O_5$ melts at this temperature. [24] A two-step heating treatment in a mildly reducing atmosphere of (99.5% Ar/0.5% $H_2$) allowed achieving the desired synthesis, with no rest of precursors or other over-reduced products.

After this calcination state, the $Fe_2VO_4$ powder was manually ground with a stone mortar and milled by ball attrition with ethanol as dispersant. A submicron median size was found for the grains after 1 hour of ball attrition process, time beyond which the particle size and the size distribution stay relatively constant. Granulometry measurements as a function of time are shown in the Supplementary Information (**Figure S1**).

The product was thoroughly washed out from the mill balls, and the resulting mixture was dried out. The fine dry $Fe_2VO_4$ powder was then mixed with a small amount of polyvinyl alcohol and uniaxially pressed under a 60-bar load to form a 4 mm thick and 15 mm diameter pellet. Finally, the pellet was sintered at 1300ºC, reached with a 100ºC/h speed, during 20 h in a mild reducing atmosphere (99.5% Ar/0.5% $H_2$) and cooled down to room temperature at 200ºC/h under the same atmosphere.

$Fe_2VO_4$ thin films (40 nm thick) were grown by PLD by ablating this sintered target with a KrF excimer laser (λ = 248 nm). Commercial MgO(001) single crystals (a = 4.211 Å, CODEX) [25] were chosen as substrates due to their small lattice mismatch (- 0.2%). The target-to-substrate distance was kept at 5 mm. Any $O_2$ pressure would result into the oxidation of V(III) into V(V) and the formation of a $V_2O_5$ sublayer. [17] Ar was thus chosen as the deposition gas. The deposition conditions were inspired from those optimized for the deposition of the $FeV_2O_4$ thin films, varying the deposition temperatures (between 400ºC and 600ºC), the Ar partial pressures (0.001 to 0.1 mbar), the laser fluences (from 1.3 to 4 J cm$^{-2}$) and the laser deposition rates (2 to 10 Hz). The optimized conditions were 400ºC, 0.01 mbar, 4 J cm$^{-2}$ and 5 Hz, with temperature and Ar pressure being the most influential parameters for the resulting film's quality. Reflection High-Energy Electron Diffraction (RHEED) measurements were carried out to verify the thin film's electronic diffraction pattern, and to carry out an *in situ* monitoring of the deposition process.

The crystal phase of the target was verified by X-ray diffraction using a D8 Brucker diffractometer in the Bragg-Brentano configuration equipped with a front monochromator, a copper anode ($K_{\alpha 1}$ = 1.54056 Å) and an energy resolved Lynxeye XE-T linear detector. The Fe/V stoichiometry in the target was verified using energy dispersive spectroscopy (EDS) coupled to a scanning electron microscope (SEM) using a *JEOL JSM* 6700F microscope equipped with a field emission gun operating at an accelerating voltage of 15 kV.

The topography of the thin films was studied with non-contact Atomic Force Microscopy (AFM) measurements with a *Park Systems* XE7 equipment. Their thickness and crystallographic structure were determined by X-Ray reflectivity (XRR) and diffraction (XRD) using *a Rigaku SmartLab* diffractometer equipped with a Ge 220 x2 front monochromator and a copper anode ($K_{\alpha 1}$ = 1.54056 Å). θ-2θ scans allowed the study of the out-of-plane lattice parameters. ϕ-scans were performed to verify the epitaxial relationships between film and substrate and reciprocal space mappings (RSM) allowed the determination of all the lattice parameters of the thin film. Resonant Elastic X-ray Scattering (REXS) experiments in the X-ray Absorption Near Edge Structure (XANES) domain were performed at the D2AM beamline of the ESRF Synchrotron Radiation Facility at the Fe K-edge to determine the cationic occupation of both Oh and Td sites. High-resolution θ-2θ scans were also obtained in the D2AM beamline for a more precise determination of the out-of-plane lattice parameter, with λ = 1.74331 Å.

The magnetic properties of the films have been characterised using a superconducting quantum interference device vibrating sample magnetometer (SQUID VSM MPMS 3, *Quantum Design*). Finally, measurements of the resistivity of the films as a function of temperature have been conducted using a cryo-free physical properties measurements system (Dynacool PPMS, *Quantum Design*). Resistivity values have been obtained using the four-contact van de Pauw method [26]–[28] by measuring the voltage under a constant current input of $1.0 \times 10^{-5}$ A. All the results shown hereafter correspond to the same optimised VFO//MgO film. The reproducibility of the deposition process was checked by analysing two other samples deposited with the same conditions.

### 3. Results and discussion
### 3.1.     $Fe_2VO_4$ ceramic target

The powder X-ray diffractogram of the VFO ceramic target is shown in **Figure 2**. A profile matching analysis using the FULLPROF program [29] confirmed the cubic spinel structure with a = 8.429(2) Å, close to the value reported in the literature for VFO: a = 8.418(2) Å. [10] EDS analyses indicated an Fe-to-V ratio of 1.98(3) – close to the expected value of 2.0.

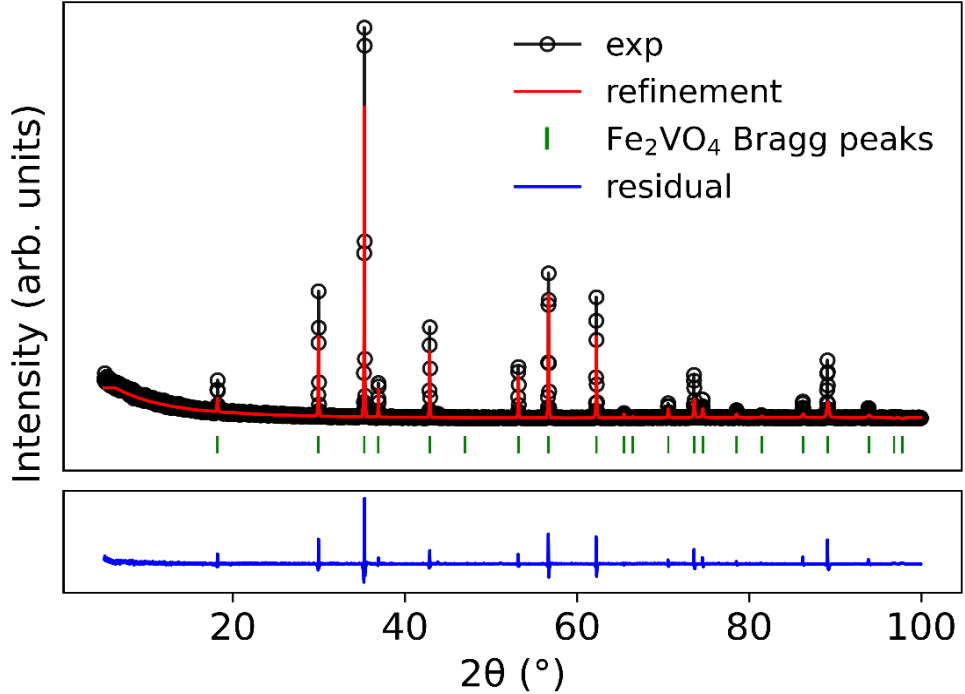

**Figure 2**. X-ray characterisation of the VFO target, showing the experimental diffractogram (black), the FULLPROF refinement (red), the position of the Bragg peaks predicted for a $Fd\bar{3}m$ structure (green) and the residual between the experimental and refined data (blue).

### 3.2. Fe$_2$VO$_4$ thin films

The AFM image of a 5 × 5 µm² surface area is shown in **Figure 3b** together with the image of a MgO substrate prior to deposition (**Figure 3a**), for comparison. The AFM images show a relatively smooth surface for both the substrate and the thin film. Root mean square roughness ($R_q$) values were obtained from region analysis studies on the shown areas and are of ≈0.5 nm for MgO and ≈0.6 nm for VFO//MgO.

The RHEED pattern observed before deposition (**Figure 3c**) consists of a series of streaks, as expected for single-crystalline MgO substrates. The images show Kikuchi lines, indicator of a high quality of the crystal surface. [30]–[32] The RHEED pattern of VFO//MgO after deposition consists of rod-shaped spots which is associated to a 2D smooth and flat surface, and still shows Kikuchi lines, indicating both a good surface state and crystallinity inside the material. [33]

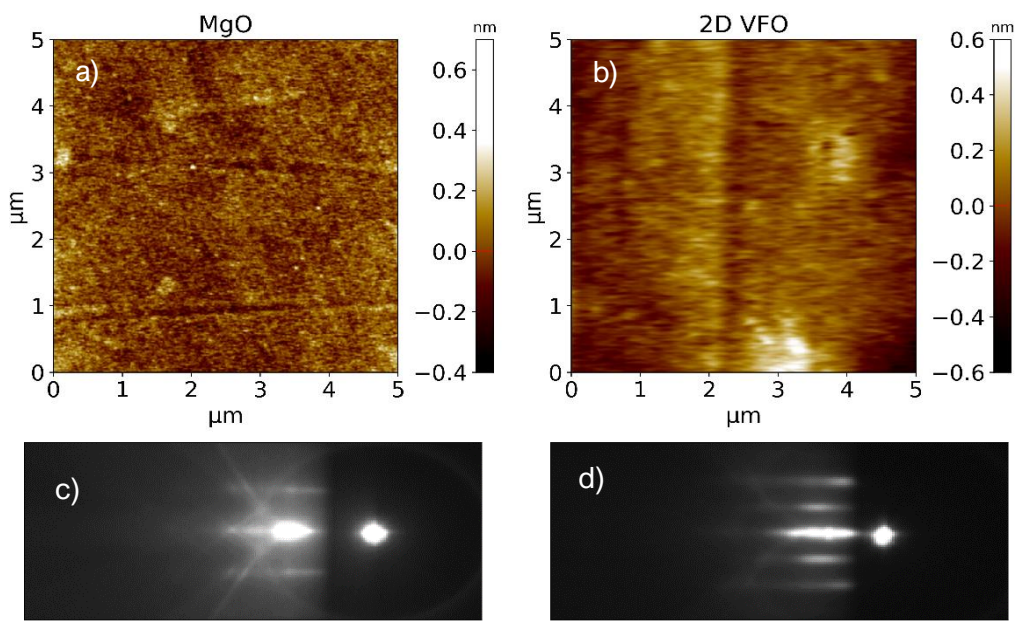

**Figure 3**. AFM images of (**a**) MgO prior to deposition and (**b**) VFO//MgO films, both acquired in non-contact mode. RHEED patterns before (**c**) and after (**d**) deposition.

XRR measurements allowed determining that the sample's thickness is 40.8(2) nm, as shown in the Supplementary Information (**Figure S2**). The film was finely characterized in its out-of-plane direction thanks to a precise θ-2θ diffractogram on the 004 reflection (**Figure 4a)** with a wavelength of 1.74331 Å. A diffractogram covering a broader θ-2θ range is shown in the Supplementary Information (**Figure S3**). Compared to the bulk sample, only the $00l$ reflections are observed, confirming the film epitaxy and the absence of parasitic phases. The diffractograms show a fringe-like profile centred around the peak's maximum. The presence of these oscillations, known as Laue oscillations, is frequently used as evidence of high sample quality, indicating that grown films are highly crystalline, homogenous, have a low density of defects and possess a smooth substrate/film interface [34], [35]. The 004 peak was fitted using the layer builder of the xrayutilities package [36]. The position of the peaks indicates an out-of-plane lattice parameter of 8.408(2) Å for VFO and 4.2113(1) Å for MgO, in perfect agreement with the 4.211 Å theoretical value [25]. The oscillation frequency was used to determine the sample's thickness: 38.7(1) nm, close to the value determined via XRR: 40.9(2) nm.

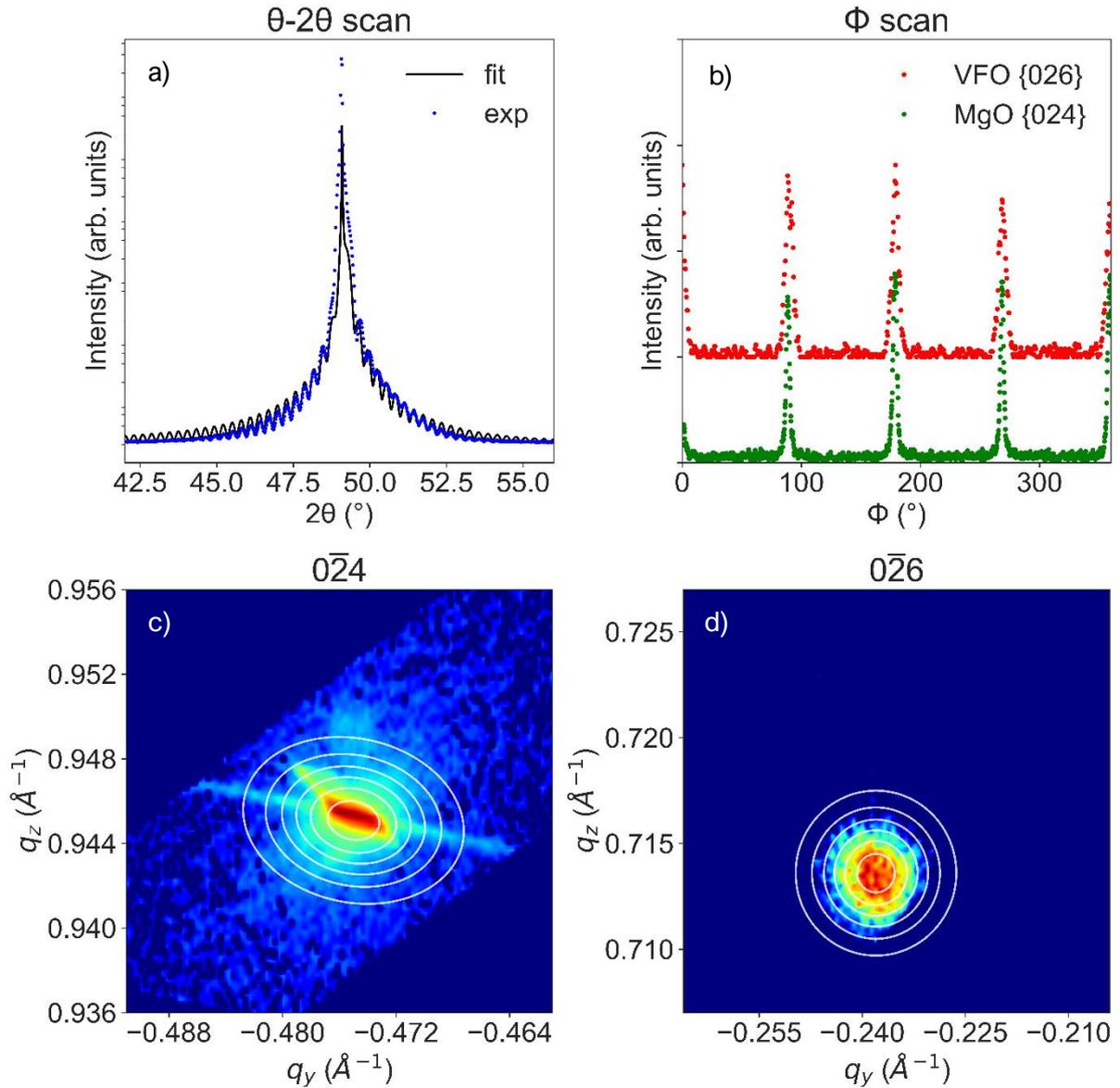

**Figure 4.** (**a**) θ-2θ diffractogram of the VFO//MgO thin films, acquired with an X-ray wavelength of 1.74331 Å – experimental data (black) and xrayutilities fit (blue). (**b**) Φ scan on a 026 reflection of VFO and 024 reflection for MgO. (**c**) reciprocal space mappings acquired for the $0\bar{2}4$ reflection of MgO (eclipsing the $0\bar{4}8$ reflection of VFO) and the $0\bar{2}6$ reflection of VFO.

Azimuthal Φ scans were carried out at constant 2θ – ω angles, or the VFO {026} and secondly for the MgO {024} reflections, as shown in **Figure 4b**. Φ = 0 represents a parallel alignment with the substrate's edge. Both scans show four distinct equidistant peaks in the 0 – 360° range. The presence of film and substrate peaks at the same angles is evidence of the crystallographic alignment between their structures, and therefore, of the epitaxial nature of the growth of the thin film system.

Finally, the 2D-RSM images probed the substrate and the film simultaneously for the MgO 024 reflection and the VFO 048 reflections located in close vicinity (**Figure 4c**), and only the film for its 026 reflection of FVO (**Figure 4d)**. The sole presence of clearly defined peaks for the films is another proof of a well-crystallised single-phased film.

The substrate's signal eclipses the film's one in the MgO 024 – VFO 048 map, since they appear at very close $q_z$ and $q_x$ values. A lattice parameter of $a$ = 4.213(2) was found for MgO, in concordance with the theoretical value of 4.211 Å [25]. The film's 026 reflection, not accompanied with any substrate reflection, enabled a determination of the lattice parameters: $a$ = $b$ = 8.415(6) Å and $c$ = 8.407(3) Å. The in-plane lattice parameter is very close to twice that of MgO: 4.211 × 2 = 8.422 Å. Consequently, the VFO lattice is compressed in a fully strained mode to adjust its lattice parameter to MgO, as expected for the epitaxial nature of the deposition and enabled by the small lattice mismatch (≈ - 0.1%). The out-of-plane lattice parameter is consistent with the value found via θ-2θ measurements. The lattice parameters of the substrate and the film have been determined by the mapping of their four equivalent {048} and {026} reflections, respectively.

Compared to their bulk counterparts, VFO thin films suffer a lattice shrinking of ≈ -0.7% and a strain-based structural tetragonalization. We observe that an in-plane lattice compression does not translate into an out-of-plane lattice elongation, as it would be expected. The simultaneous shrinking of both lattice parameters corresponds to a mechanical response known as "auxetic behaviour". Unlike conventional materials, which expand in one direction when compressed in another, auxetic materials exhibit a counterintuitive phenomenon where they can compress in both directions when submitted to a compression in one direction. Auxetism in oxide materials is rare but it has also been reported in other spinel thin film systems such as $CoFe_2O_4$//MgO [37], $NiFe_2O_4$//STO [38] or $FeV_2O_4$//MgO [17]. Auxetism can be also quantified through the negative value of the effective Poisson's ratio [39] $v^* \approx$ - 0.8, whose calculation for epitaxial thin films with biaxial (symmetric) in-plane stress is described in **Equation (2)**. Its value depends on $v$, the Poisson coefficient $v = - \epsilon_t/\epsilon_l$ which is the ratio between the strain in the transverse ($\epsilon_l = (a_{film} - a_{bulk})/a_{bulk}$) and the longitudinal ($\epsilon_l = (c_{film} - a_{bulk})/a_{bulk}$) directions. [40]

$$v^* = \frac{2v}{1 - v} \approx -0.8 \quad \textbf{(2)}$$

Finally, EDS measurements on the thin films led to an Fe/V ratio of 2.2(3), slightly higher than the expected value of 2.0. In the soft X-ray region, the vanadium signal suffers significant interference from oxygen [41] due to the close proximity of O K-edge (530 eV) [42] with the V

L-edges (520 – 530 eV) [43], resulting in an underestimation of the vanadium content and an overestimation of the Fe/V ratio. For higher X-ray energies, the beam's penetration length is larger than the thin film thickness, and the elements which compose the substrate crystal (Mg, O) are unavoidably included in the quantification and their presence in the film is also overestimated. The substrate also presents a non-negligible amount of Fe which furtherly increases the Fe/V ratio.

### 3.3. Cationic distribution of VFO films

REXS experiments were performed to determine the cationic distribution of the Fe and V species in the Oh and Td sites. We have characterised the positions of both Fe and V cations by studying XANES (x-ray absorption near the edge structure) spectra acquired at the Fe K-edge – around 7130 eV [44]. REXS simultaneously allows an element and orbital-sensitive analysis of the crystal [45], and the XANES region of the spectrum can give fruitful insights on the cation's state. REXS-XANES (also known as DANES: diffraction anomalous near-edge structure) has already been proved successful in characterising the cation's state (occupation and valence state) in thin films. [46], [47] Furthermore, we have previously used REXS-EXAFS experiments to successfully locate oxygen cations in oxide thin films. [48]

By using the *inserexs* software, as well as the methodology that we have developed along the code [49], we have chosen the reflections 333 and 115 as the most adequate for the characterisation of the cations. REXS spectra (-20 eV, +25 eV) have been acquired around the Fe K-edge (- 20 eV, + 25 eV) for both reflections. This edge choice, compared to the V K-edge, is dictated by the smaller wavelength of the incident X-rays which results in fewer self-absorption issues and a better signal-to-noise ratio. [48]

Theoretical spectra with the formula $(Fe_{1-x}V_x)_{Td}(V_{2(1-y)}Fe_{2y})_{Oh}O_4$ have been simulated with $x$ and $y$ values varying between 0 and 1 with increments of 0.005. The simulations have been conducted using the *ab initio* FDMNES software [50] and the simulation conditions have been detailed in the Supplementary Information. The difference between the experimental ($ES$) and the theoretical ($TS$) spectra ($\chi^2 = \int TS(E) - ES(E) dE$) has been calculated for all ($x$, $y$) combinations, as illustrated in **Figure 5**. $\chi^2$ values vary with both $x$ and $y$, proving that the technique is valid to probe the cationic species at both Td and Oh sites. The $\chi^2$ minimum (best fit) unequivocally takes place at $x$ = 0.021(8) and $y$ = 0.54(2), as it appears in **Figure 5a**. The spectra simulated at these conditions perfectly fit the experimental data acquired at the 333 and 115 reflections, as indicated in **Figure 5b** and **5c**, respectively.

This ($x$, $y$) minimum leads to a cationic occupation of $(Fe_{0.979(8)}V_{0.021(8)})_{Td}[V_{0.92(4)}Fe_{1.08(4)}]_{Oh}O_4$, that is, an experimental formula unit of $V_{0.94(5)}Fe_{2.05(5)}O_4$ for VFO films and an Fe/V ratio of 2.1(1), consistent with the ratio observed in the bulk of 1.98(3). These values are in adequacy with the EDS results and are a confirmation of the stoichiometric transfer of matter from the target to the film during deposition. The cationic occupation is very close to the expected scenario (Fe)[VFe]$O_4$ in which all the V ions occupy Oh sites, whereas the Fe ones occupy both the Td and Oh sites.

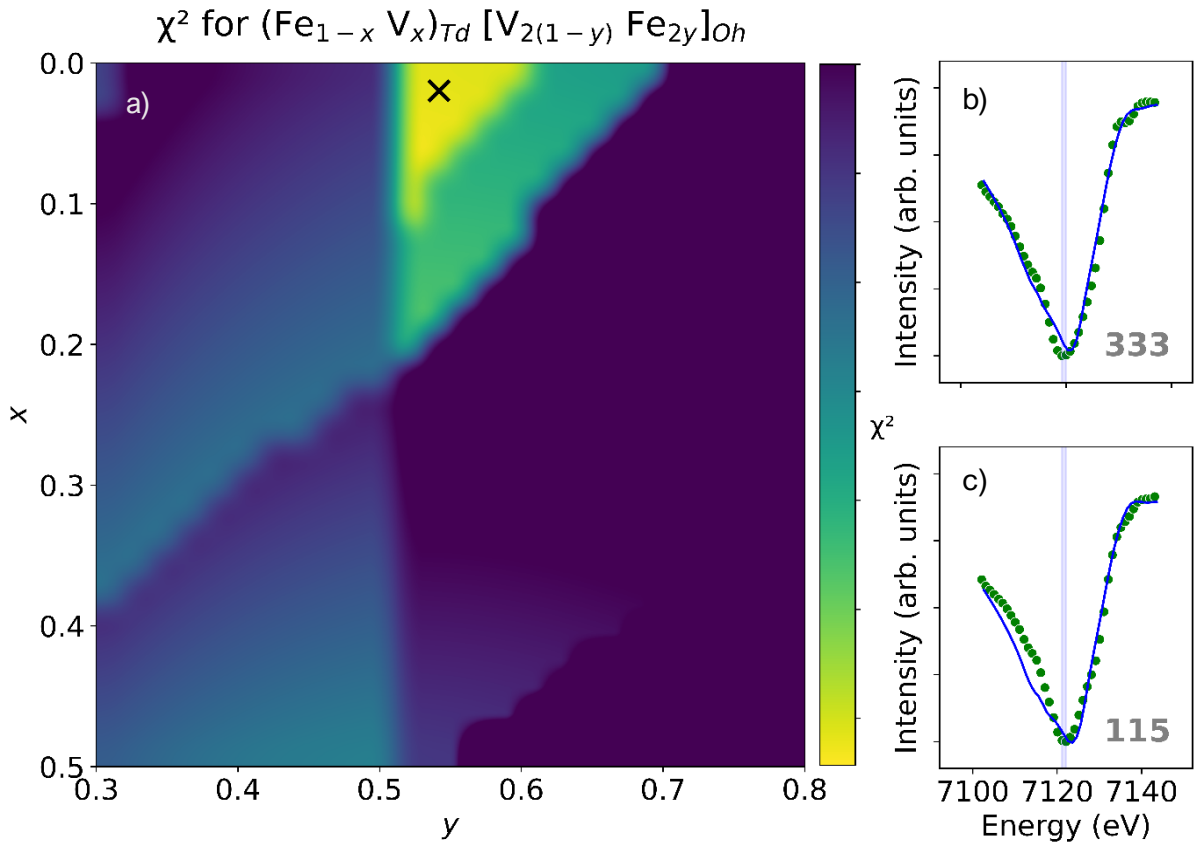

**Figure 5**. (**a**) Difference calculated between the experimental spectra and the theoretical spectra simulated for all ($x$, $y$) combinations. (**b**) and (**c**) experimental spectra for the 333 (b) and 115 (c) reflections (green points) and fit with the spectrum simulated for the best ($x$, $y$) (blue line)

### 3.4. Magnetic and transport properties of VFO films

The magnetic response of the thin films was characterised via magnetisation (*M*) vs. applied field (*H*) measurements at 300 K in in-plane (IP) and out-of-plane (OOP) configurations, as shown in **Figure 6a**. The sample shows a non-negligeable saturation magnetization ($M_s$) of around 0.8 $\mu_B$/f.u. with an easy magnetization axis in the in-plane direction and an IP coercive field of 205(5) Oe (1.98(4) × $10^4$ A $m^{-1}$). The $M_s$ value is consistent with the one already reported for the bulk. [10] The sample therefore presents a magnetic ordering at room temperature. No magnetic transition temperature has been detected in the temperature range 2 – 400 K, as shown in the Supplementary Information (**Figure S4**). However, the decrease of the magnetic signal when approaching 400 K tends to indicate that the ordering temperature is close to that temperature, and matches the transition temperature reported for bulk VFO of 454 K. [10]

The transport properties of the films have been characterised via sheet resistivity vs. temperature measurements in the range 250 – 400 K, as illustrated in **Figure 6b**. Voltage values exceeded the limit value of our PPMS below 250 K. The temperature evolution of the the material's resistivity can be modelled by a thermal activation model: $\rho = \rho_0 e^{-\frac{E_a}{kT}}$, where $\rho_0$ is a pre-exponential factor - theoretical resistivity at infinite temperature, $E_a$ is the activation energy for the thermal activation of electrons and $k$ is the Boltzmann's constant. The resistivity thus follows an Arrhenius-like activation process of charge carriers whose presence in the conduction band becomes more notorious with increasing temperature. [51] A discontinuity in the model trend starts to appear at around 400 K, also suggesting the vicinity of a phase transition.

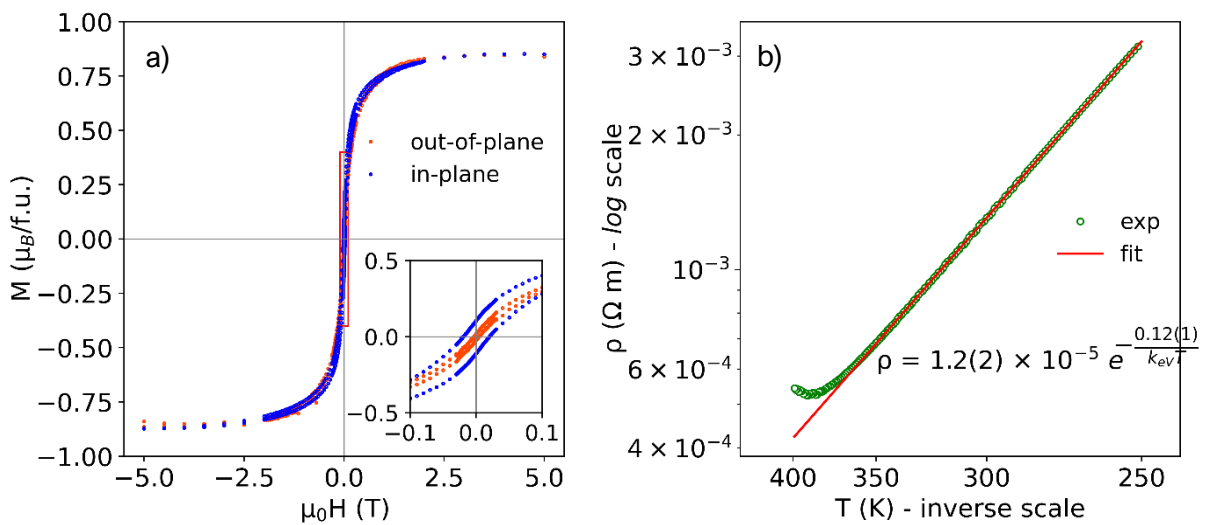

**Figure 6**. a) Magnetisation hysteresis loops measured at 300 K, for the in-plane (blue) and out-of-plane (red) configurations. b) Longitudinal resistivity (*log* scale) vs. temperature ($^{-1}$ scale) – experimental data in green circles, fit using the thermal activation model in solid red line.

Using the elementary distribution determined by the DANES analysis, and introducing the valence of the cations, the cation distribution can vary between a ($Fe^{2+}$)[$Fe^{3+}V^{3+}$]$O_4$ direct spinel configuration and a $(Fe^{3+})[Fe^{2+}V^{3+}]O_4$ inverse spinel one. In both direct and inverse spinel configurations, an antiparallel alignment of the magnetic moments between cations at Oh and Td sites is expected [52], [53], and the observed moment of $\mu_{VFO} = 0.8\ \mu_B/f.u.$ can therefore be written as $\mu_{VFO} = |\mu_{Oh} - \mu_{Td}|$. Considering that the magnetic contribution of both $Fe^{3+}$ and $Fe^{2+}$ ions comes solely from the spin-only contribution ($\mu_{Fe^{2+}} \approx 4\mu_B$, $\mu_{Fe^{3+}} \approx 5\mu_B$), and that $V^{3+}$ will align along the species in the Oh site, one can calculate the magnetic contribution of $V^{3+}$. The net magnetic contribution of $V^{3+}$ is expected to be smaller than a spin-only scenario ($\mu_{V^{3+}-SO} \approx 2\mu_B$) due to both canting phenomena, as reported in other spinel vanadates [54], and to the quenching of magnetic moments due to the spin-orbit coupling [11].

Two possible scenarios for the magnetic moment of $V^{3+}$ arise depending on whether VFO adopts a direct ($\mu_{V^{3+}} = \mu_{Fe^{2+}} - \mu_{Fe^{3+}} \mp \mu_{VFO}$) or an inverse ($\mu_{V^{3+}} = \mu_{Fe^{3+}} - \mu_{Fe^{2+}} \mp \mu_{VFO}$) spinel structure. For the direct spinel structure, the ≈0.8 $\mu_B$ per unit cell cannot be reached with a parallel alignment between the $V^{3+}$ and $Fe^{3+}$ cations in Oh sites. The only possible combination is found for the inverse spinel configuration, with $V^{3+}$ contributions of either ≈0.2 $\mu_B$ or ≈1.8 $\mu_B$ that accompany $Fe^{2+}$ in the Oh site. The latter value is the most probable since it is both closer to a spin-only scenario and to the experimental values of the orbitally quenched magnetic moment for $V^{3+}$ in FVO. [11] As a result, VFO is expected to adopt an inverse spinel structure. These findings are consistent with the neutron diffraction studies carried out in bulk VFO [9] and can be explained by the more stable configuration of $[Fe^{2+}]_{Oh}$ when compared to its trivalent counterpart.

## 4. Conclusion

The study hereby presented provides substantial insights into the synthesis, structural characterisation, and physical properties of VFO thin films, never reported yet. The thin films show a 2D grow, with a very low top roughness ($R_q \approx 0.6$ nm) and high crystalline quality. The nature of the VFO crystallographic phase was determined by XRD measurements to be a spinel one, growing epitaxially onto the MgO(001) substrate with the following lattice parameters: $a = b = 8.415(6)$ Å and $c = 8.407(3)$ Å. The spinel phase is tetragonally distorted because of the compressive strain effects, and the films exhibit an auxetic behaviour.

The stoichiometry of the VFO films have been confirmed by EDS and REXS measurements. The films are magnetically ordered at room temperature with a non negligible saturation magnetisation of around 0.8 $\mu_B$/f.u. The cationic distribution could be determined combining magnetic and REXS measurements. VFO adopts an inverse spinel structure close to ($Fe^{3+}$)[$Fe^{2+}V^{3+}$]$O_4$. Both the magnetic and electric properties of the films suggest a possible phase transition around 400 K. The resistivity below this temperature can be modelled using a thermal activation model.

In conclusion, our findings open the path of potential applications of $Fe_2VO_4$ thin films in new room temperature electronic devices.

**Acknowledgements**

This work of the Interdisciplinary Thematic Institute QMat, as part of the ITI 2021 2028 program of the University of Strasbourg, CNRS and Inserm, was supported by IdEx Unistra (ANR 10 IDEX 0002), and by SFRI STRAT'US project (ANR 20 SFRI 0012) and EUR QMAT ANR-17-EURE-0024 under the framework of the French Investments for the Future Program. The authors acknowledge the collaboration of the 'Meb-Cro' and 'DRX' platforms of the IPCMS.

**Competing interests**

The authors declare no conflict of interest.

## 2. Supplementary information

### 1. Granulometry

During the ball attrition milling process, supernatant samples of a few millilitres were analysed with a granulometer at different attrition times, as shown in **Figure S1.** The size distributions (**Figure S1a**) have been fitted using a Gaussian distribution function: $f(x) = \frac{1}{\sigma\sqrt{2\pi}} e^{-(x-\mu)^2/(2\sigma^2)}$. The particle size has been described by the centre of the Gaussian distribution ($\mu$) or expectation, whereas the amplitude of the size distribution has been characterised by the standard deviation ($\sigma$). Just before the ball milling process, particles have a size of around 5 µm. Then, the attrition process starts breaking particles down, but they seem to agglomerate into bigger particle clusters. That drives both the mean size and the standard deviation to increase. At a certain time, the particles are broken down into stable sizes, and both $\mu$ and $\sigma$ take a low value. Beyond a 60 minutes attrition time, the particles reach a submicron size with a narrow size distribution. The evolution of both $\mu$ and $\sigma$ as a function of time has been plotted in **Figure S1b**.

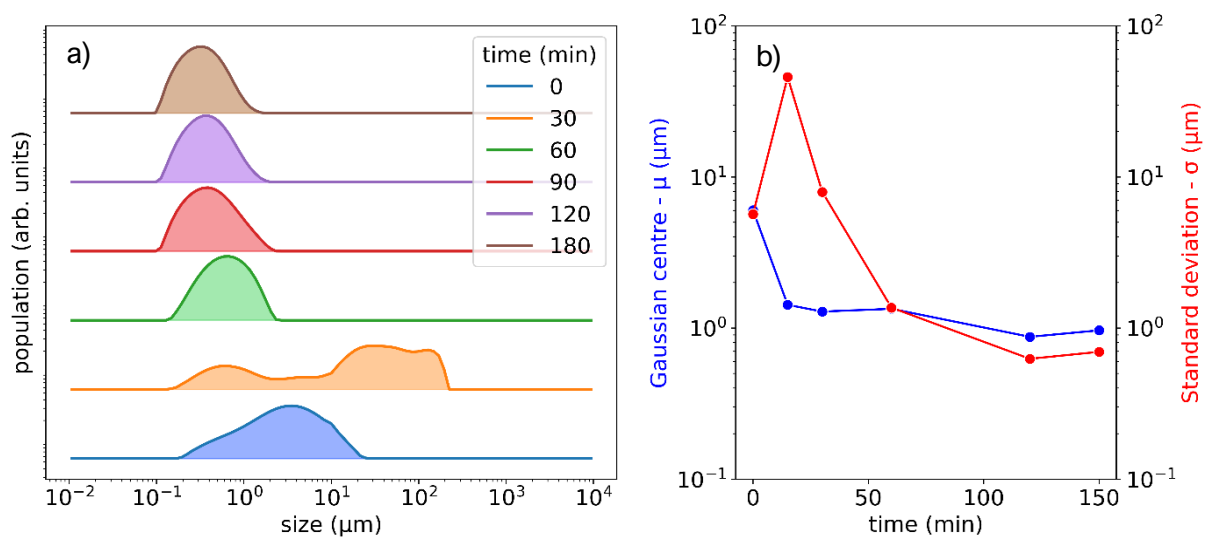

**Figure S1.** (**a**) Population distributions as a function of size for attrition times ranging from 0 to 180 minutes. (**b**) Centre of gaussian peaks for the population distribution (blue) and standard deviations of the distributions (red) as a function of time.

### 2. X-ray reflectivity measurements

XRR measurements allowed a precise determination of the film's thickness: 40.8(1) nm, along with other parameters such as the film's roughness: $R_q$ = 0.9(1) nm, the interface roughness: $R_q$ = 1.0(1) nm and the film's density: 5.0(1) g cm$^{-3}$, as shown in **Figure S2**.

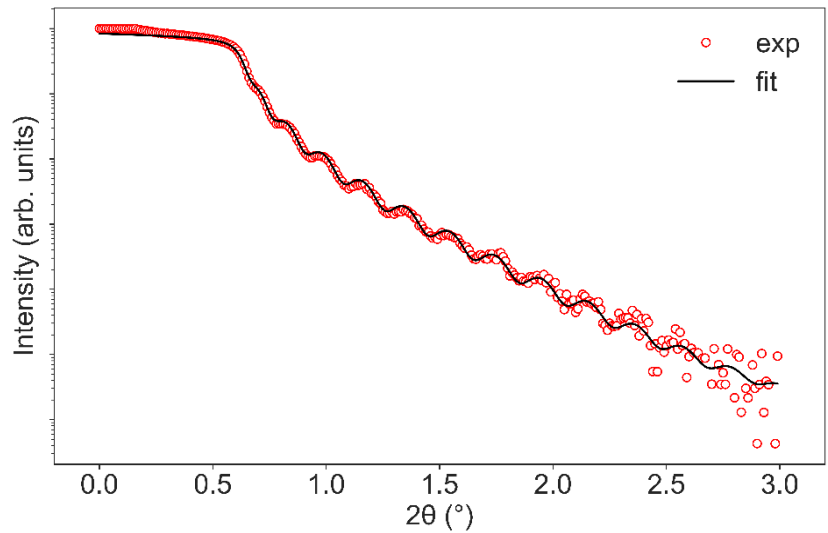

**Figure S2**. XRR – measurement (red) and refinement (black) for a VFO//MgO thin film.

This higher roughness value, compared to the one obtained via AFM measurements (≈0.6 nm) is due to the short wavelength of probing X-rays, the convolution of the AFM tip shape and the fact that XRR can probe buried interfaces beneath the top surface layer, which is beyond the reach of AFM. [55]

3. **Large range θ-2θ diffractogram**

θ-2θ scans were carried out in the [30° - 110°] range, as shown in **Figure S3**. No parasitic crystalline phases were detected. Only the $00l$ peaks for VFO (004 and 008) and MgO (002 and 004) were observed, which is a further proof of the film-to-substrate epitaxy. Laue oscillations are still observed for both VFO peaks, just as shown on **Figure 4a** for the 004 peak.

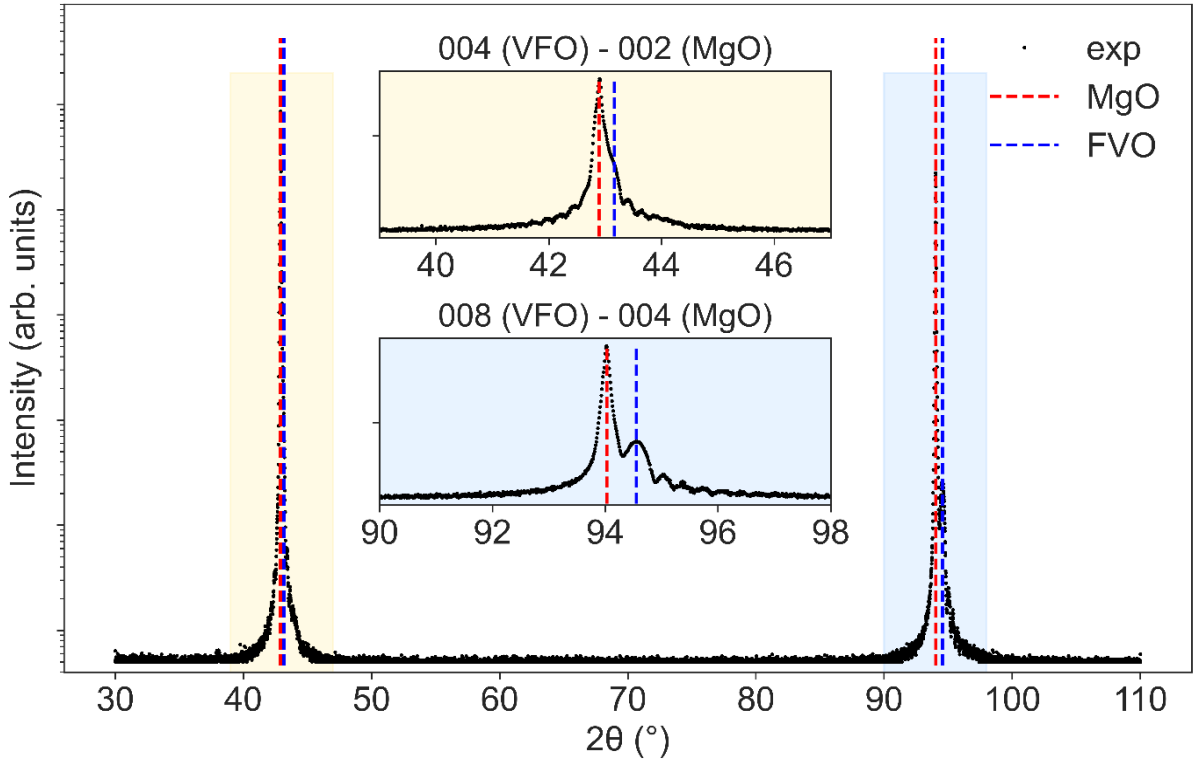

**Figure S3**. θ-2θ diffractogram for the study of the $00l$ reflections of MgO (red dashed lines) and VFO (blue).

## 4. Simulations conditions for FDMNES

Simulations of spectra have been conducted within the energy range of [-20, +25] around the absorption Fe K-edge with a 1.0 energy step, following the experimental conditions for the spectra acquisitions.

A calculation radius of 5 Å was selected, which appeared to be an appropriate choice relative to the lattice parameter of the system, which is around 8 Å. Larger radius values yielded negligible variation in the spectra's shape but significantly escalated the simulation time. On the contrary, smaller radius values led to substantial alterations in the spectra.

For the simulations, the atomic coordinates were kept constant for the cations, which are positioned at fixed special locations. Lattice parameters for the spectra simulations have been taken from reciprocal space mapping (RSM) analyses. Minimal to no alterations in the final outcome result when slightly modifying the lattice parameters, so the effects of their incertitude has been neglected.

## 5. Magnetization vs. temperature

The magnetization as a function of temperature has been measured in the [2 - 400 K] range under an applied magnetic field of 0.5 T, as shown in **Figure S4**. The temperature boundaries correspond to the measurement limits of our PPMS.

The raw data (**Figure S4a**) shows a predominant paramagnetism, due to the presence of Fe impurities in the MgO substrate (confirmed by EDS). Similar measurements on a virgin MgO have confirmed the paramagnetic behaviour of the substrate. As mentioned in the main text and after paramagnetic correction (**Figure S4b**, by fitting with Curie's law and subtracting the paramagnetic component), no magnetic transition temperature has been detected in the aforementioned temperature range. Nevertheless, the decrease of the magnetic signal when approaching 400 K tends to indicate that the ordering temperature is close to that temperature, and matches the transition temperature reported for bulk VFO: 454 K. [8]

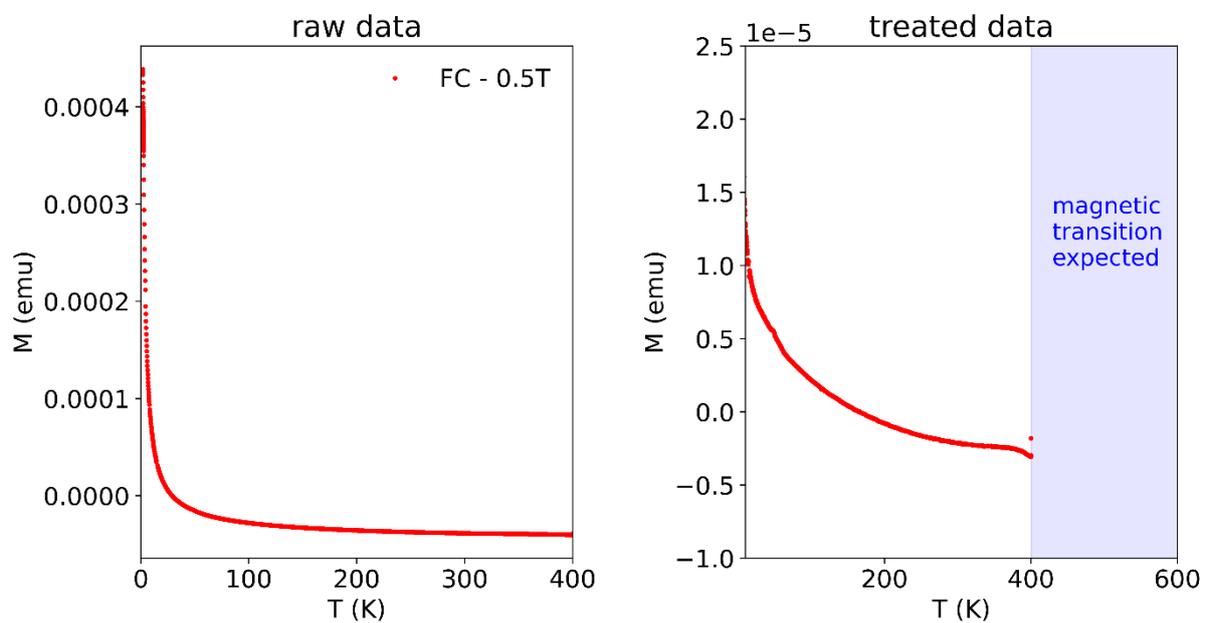

**Figure S4**. Magnetization as a function of temperature: (**a**) raw data, (**b**) after paramagnetic correction